\documentclass[11pt]{article}
\usepackage{amssymb}
\usepackage{graphicx}
\usepackage{lscape}
\usepackage{theorem}
[1999/03/29 PSNFSS v.7.2
Times font as default roman
: S Rahtz]

\newcommand \be  {\begin{equation}}
\newcommand \bea {\begin{eqnarray}}
\newcommand \ee  {\end{equation}}
\newcommand \eea {\end{eqnarray}}
\newcommand \E {{\rm E}}

\newcommand \w {{\omega}}
\newcommand \Cov {{\rm Cov}}
\newcommand \Var {{\rm Var}}
\newcommand \Dt {{\Delta t}}

\begin{document}

\title{Endogeneous Versus Exogeneous Shocks in Systems with Memory
\footnote{We acknowledge helpful discussions and
exchanges with Y. Malevergne, J.-F Muzy and V. Pisarenko. This work was partially supported by
the James S. Mc Donnell Foundation 21st century scientist award/studying
complex system.}}
\author{D. Sornette$^{1,2}$ and A. Helmstetter$^3$\\
\\
$^1$ Institute of Geophysics and Planetary Physics
and Department of Earth and Space Science\\
University of California, Los Angeles, California 90095, USA\\
$^2$ Laboratoire de Physique de la Mati\`ere Condens\'ee CNRS UMR 6622\\
Universit\'e de Nice-Sophia Antipolis, 06108 Nice Cedex 2, France\\
$^3$ Laboratoire de G{\'e}ophysique Interne et Tectonophysique\\
Observatoire de Grenoble, Universit{\'e} Joseph Fourier, BP 53X\\
38041 Grenoble Cedex, France\\
email: sornette@ess.ucla.edu  and  ahelmste@obs.ujf-grenoble.fr\\
fax:  (310) 206 30 51\\
}

\maketitle

\begin{abstract}

Systems with long-range persistence and memory are shown to 
exhibit different precursory as well as recovery patterns 
in response to shocks of exogeneous versus endogeneous origins.
By endogeneous, we envision either fluctuations resulting
from an underlying chaotic dynamics or from a stochastic forcing origin
which may be external or be an effective coarse-grained description
of the microscopic fluctuations.
In this scenario, endogeneous shocks result from a kind of 
constructive interference of
accumulated fluctuations whose impacts survive longer than the
large shocks themselves. As a consequence, the recovery after
an endogeneous shock is in general slower at early times and can be
at long times
either slower or faster than after an exogeneous perturbation.
This offers the tantalizing possibility of distinguishing between
an endogeneous versus exogeneous cause of a given shock, even
when there is no ``smoking gun.'' This could help in 
investigating the exogeneous versus self-organized origins in problems
such as the causes of major biological extinctions, of 
change of weather regimes and of the climate,
in tracing the source of social upheaval and wars, and so on.
Ref.~\cite{Soretalfingerfin} has already shown how this concept can be
applied concretely to differentiate
the effects on financial markets of the
Sept. 11, 2001 attack or of the coup against Gorbachev on
Aug., 19, 1991 (exogeneous) from financial 
crashes such as Oct. 1987 (endogeneous).

\end{abstract}

\vskip 1cm

\section{Introduction}

Most complex systems around us
exhibit rare and sudden transitions, that occur over
time intervals that are short compared to the
characteristic time scales of their posterior evolution.
Such extreme events express more than anything else the underlying ``forces'' usually
hidden by almost perfect balance and thus provide the potential for a better
scientific understanding of complex systems. 
These crises have fundamental societal
impacts and range from large natural catastrophes such as earthquakes, volcanic
eruptions, hurricanes and tornadoes, landslides, avalanches, lightning strikes,
meteorite/asteroid impacts, catastrophic events of environmental degradation, to
the failure of engineering structures, crashes in the stock market, social unrest
leading to large-scale strikes and upheaval, economic drawdowns on national and
global scales, regional power blackouts, traffic gridlock, diseases and
epidemics, and so on. It is essential to realize that
the long-term behavior of these complex systems is
often controlled in large part by these rare catastrophic events
\cite{Physicsworld}.
The outstanding scientific question is how such large-scale patterns
of catastrophic nature might evolve from a series of interactions on the smallest
and increasingly larger scales \cite{Predict}, or whether their origin should be searched
from exogeneous sources.

Starting with Hurst's study of 690 time series records of 75 
geophysical phenomena, in particular river flow statistics, documenting
the so-called ``Hurst effect'' of long term persistence \cite{Hurst1}, 
many studies in the last decades have investigated the existence of
long memory effects in a large variety of systems, including
meteorology (wind velocity, moisture transfer in the atmosphere, precipitation), 
oceanography (for instance wave-height),
plasma turbulence, solar activity, stratosphere chemistry, seismic activity,
internet traffic, financial price volatility, cardiac activity, immune
response, and so on. 

The question addressed here is whether the existence of long memory
processes may lead to specific signatures in the precursory and in the
relaxation/recovery/adaptation of a system after a large fluctuation of its
activity, after a profound shock or even after a catastrophic event, that may allow one
to distinguish an internal origin from
an exogeneous source. Let us put this question in perspective
with regards to the extinction
of biological species as documented in the fossil record. During the past 550 million
years, there have been purportedly five global mass extinctions, each of which
had a profound effect on life on Earth. The last end-Cretaceous mass
extinction (with the disappearance
of $39-47\%$ of fossilizable genera and perhaps 75\% of species)  
marking the Cretaceous/Tertiary (K/T) boundary about 65 millions ago is often attributed
to the impact of a huge meteor in the Yucatan peninsula \cite{Kyte}.
Another scenario is that a burst of active volcanism was the real
origin of the extinction \cite{Courtillot,Court2}. It has been 
suggested that this extinction was actually driven by
longer-term climatic changes, as evidence by the fact that
certain species in the Late Maastrichtian
disappeared a distinctive time before the K/T boundary
\cite{Marshall1,Marshall2}. A completely
endogeneous origin has also been proposed, based on the concepts
of nonlinear feedbacks between species \cite{clemens,Allen} illustrated by
self-organized criticality and punctuated equilibria
\cite{Baknaturework,Soleetal} (see \cite{Kirchner} for a rebuttal). 
The situation is even murkier for the extinctions
going further in the past, for which the smoking guns, if any, are not observable
(see however the strong correlation between extinctions and volcanic traps
presented in \cite{Court2}).
How can we distinguish between an exogeneous origin (meteorite, volcanism, abrupt climate
change) and endogeneous dynamics, here defined as the progressive 
self-organizing response of the
network of interacting species that may generate its own demise by nonlinear
intermittent negative feedbacks or in response to the accumulation of 
slowly varying perturbations in the environment? Is it possible 
to distinguish two different exogeneous origins, one occurring 
over a very short time interval (meteorite) and the other
extending over a long period of time (volcanism), based on the observations
of the recovery and future evolution of diversity?

The aviation industry provides another vivid illustration of 
the question on the endogeneous versus
exogeneous origin of a crisis. 
Recently, airlines became the prime industry victim of the September 11, 2001
terrorist attacks. The impact of the downturn in air travel
has been severe not just on the airlines but also on lessors and
aircraft manufacturers. The unprecedented drop in air travel and airline performance
prompted the US government to provide \$5 billion in
compensation and to make available \$10 billion in loan guarantees. 
This seems a clear-cut case for an exogeneous shock. However,
the industry was deteriorating before the shock of September 11.
In the first eight months of 2001, passenger traffic for US
carriers rose by an anemic 0.7 percent, a sharp fall from annual
growth of nearly 4 percent over the previous decade \cite{Aviation},
illustrated by the record levels of the earned profits of \$39 billion 
and of delivery of more than 4,700 jetliners from 1995 to 2000.
The US airlines' net profits dropped from margins
of nearly 4 percent during 1998-2000 to losses of greater than 3
percent during the first half of 2001, 
despite aggressive price cuts as airlines tried to fill seats and
profits vanished. 

Many other examples are available.
We propose to address this general question of exogeneous versus
endogeneous origins of shocks by quantifying how the dynamics
of the system may differ in its response to an exogeneous versus endogeneous shock.
We start with a simple ``mean field'' model of the 
activity $A(t)$ of a system at time $t$, viewed as the effective
response to all past perturbations embodied by some noisy function $\eta(\tau)$,
\be
A(t) = \int_{-\infty}^t d\tau~\eta(\tau) ~K(t-\tau)~,
\label{mbnall}
\ee
where $K(t-\tau)$ can be called the memory kernel, propagator, Green function, or
response function of the system at a later time $t$ to a perturbation
$\eta(\tau)$ that occurred at an earlier time $\tau$. Notwithstanding
the linear structure of (\ref{mbnall}), we do not restrict
our description to linear systems but take (\ref{mbnall}) as an effective
coarse-grained description of possible complex nonlinear dynamics. For instance,
it has been shown \cite{Tanguy} that the extremal nonlinear
dynamics of the Bak and Sneppen model and of the Sneppen model of extremal evolution of
species, which exhibit
a certain class of self-organized critical behavior \cite{Sorbook},
can be accurately characterized by the stochastic process called
``Linear fractional stable motion,'' which has exactly the form (\ref{mbnall}) for
the activity dynamics.

Expression (\ref{mbnall}) contains for instance the 
fractional Brownian motion (fBm) model introduced
by Mandelbrot and Van Ness \cite{Mandelwallis1} as a simple extension 
of the memoryless random walk to account for the Hurst effect.
From an initial value $B_H(0)$, we recall that the fBm is defined by
\be
B_H(t)-B_H(0) = {1 \over \Gamma (H+(1/2))} \int_{-\infty}^t d\tau~\eta(\tau) ~K(t-\tau)~,
\label{mbnasdll}
\ee
where $d\tau~\eta(\tau)=dW_{\tau}$ is usually taken as
the increment of the standard random walk with white noise
spectrum and Gaussian distribution with variance ${\rm E}[dW_{\tau}]=d\tau$ and
the memory kernel $K(t-\tau)$ is given by
\bea
K(t-\tau) &=& (t-\tau)^{H-{1 \over 2}}~,~~~~~{\rm for}~0 \leq \tau \leq t \\
		&=& (t-\tau)^{H-{1 \over 2}} - (-\tau)^{H-{1 \over 2}}~,~~~~~~
		{\rm for}~\tau < 0~.
\eea
For $H>1/2$, the fBm $B_H(t)$ exhibits long term persistence and memory,
since the effect of past innovations of $dW_{\tau}$ is felt in the future
with a slowly decaying power law weight $K(t-\tau)$. For our purpose, the fBm is non-stationary
and it is more relevant to consider globally statistically stationary processes.

Here, we consider processes which can be described by an
integral equation of the form (\ref{mbnall}) and 
(\ref{mbnasdll}) but with possibly different
forms for the noise innovations $\eta$ and for the memory kernel $K$.
Simple viscous systems correspond to $K(t-\tau) \propto
\exp [-(t-\tau)/T]$, where $T$ is a characteristic relaxation time. 
Complex fluids, glasses, porous media, semiconductors, and so on, are
characterized by a memory kernel $K(t-\tau) \propto e^{-a(t-\tau)^{\beta}}$, 
with $0 < \beta < 1$, a law known under the name Kohlrausch--Williams--Watts law
\cite{Phillips}. It is also interesting to consider fractional noise motion
(fNm) defined as the time derivative of $B_H(t)$, which
does possess the property of statistical stationarity. A fNm is defined by (\ref{mbnall})
with 
\be
K_{\rm fNm}(t-\tau) = {1 \over (t-\tau)^{{3 \over 2}-H}} = {1 \over (t-\tau)^{1-\theta}}~,
\label{ngjlk}
\ee
for $H=1/2+\theta$. Persistence $1/2<H<1$ (respectively antipersistence $0<H<1/2$) 
corresponds to $0<\theta<1/2$ (respectively $-1/2<\theta<0$). Such a memory kernel
describes also the renormalized Omori's law for earthquake aftershocks
\cite{SorSor,Helms}.

\section{Exogeneous versus endogeneous shock}

In the following, we consider systems described by a long memory
integral (\ref{mbnall}) with kernel $K(t)$ decaying faster than $1/\sqrt{t}$
at large times, so as to ensure the condition of
statistical stationarity. This excludes the fBm which are  non-stationary processes but
includes the fNm.

\subsection{Exogeneous shock}

An external shock occurring at $t=0$ can be modeled in this framework by an innovation
which takes the form of a jump $A_0 ~\delta(\tau)$. The response
of the system for $t>0$ is then
\be
A(t) = A_0 ~K(t) + \int_{-\infty}^t d\tau~\eta(\tau)~K(t-\tau).
\label{nbghlqa}
\ee
The expectation of the response to an exogeneous shock is thus
\be
{\rm E}_{\rm exo}[A(t)] =  A_0 ~K(t) + n \langle \eta \rangle~,
\label{nmgjwslw}
\ee
where $\langle \eta \rangle$ is the average noise level and $n= 
\int_0^{+\infty} d\tau~K(\tau)$ is the average impact of a perturbation which
is usually smaller than $1$ to ensure stationarity (this corresponds
to the sub-critical regime of branching processes \cite{Harris}).

The time evolution of the system after the shock is thus the sum
of the process it would have followed in absence of shock and of the
kernel $K(t)$. The response $A_0 ~K(t)$ to the jump $A_0 ~\delta(\tau)$
examplifies that 
$K(t)$ is the Green function or propagator of 
the coarse-grained equation of motions of the system. Expression (\ref{nbghlqa})
simply expresses that the recovery of the system to an external shock
is entirely controled by its relaxation kernel.

\subsection{Endogeneous shock}

\subsubsection{Conditional response function}

Let us consider the natural evolution of the system, without any
large external shock,
which nevertheless exhibits a large burst $A(t=0)=A_0$  at $t=0$.
From the definition (\ref{mbnall}),
it is clear that a large ``endogeneous'' shock requires a special set of
realization of the innovations $\{\eta(t)\}$. To quantify the
response in such case, we recall a standard result of
stochastic processes with finite variance and covariance that the
expectation of some process $X(t)$ conditioned on some 
variable $Y$ taking a specific value $Y_0$ is given by \cite{Shiryaev}
\be
{\rm E}[X(t)|Y=A_0] - {\rm E}[X(t)] =
\left(A_0 - {\rm E}[Y]\right)~{{\rm Cov}(X(t),Y) \over {\rm E}[Y^2]}~,
\label{mgmms}
\ee
where ${\rm E}[Y^2]$ denotes the expectation of $Y^2$, ${\rm
Cov}(X(t),Y)$ is the covariance of $X$ and $Y$, ${\rm E}[X(t)]$
and ${\rm E}[Y]$ are the (unconditional) average of $X(t)$ and of $Y$.
Expression (\ref{mgmms})
recovers the obvious result that ${\rm E}[X(t)|Y=A_0]={\rm E}[X(t)]$
if $X$ and $Y$ are uncorrelated. A result generalizing (\ref{mgmms})
holds when $\eta(\tau)$ has an
infinite variance corresponding to a distribution with 
a power law tail with exponent smaller than $2$ \cite{HelmsinvOmori}.

Let us assume that the process $A(t)$ and the innovations $\eta$'s have
been defined with zero mean, which is always possible without loss of
generality by a translation. Let us call $X(t>0) = A(t)$ and $Y=A(0)$.
Under the assumption that the noise $\eta(\tau)$ has a finite variance,
we obtain from (\ref{mbnall})
\be
{\rm Cov}(A(t),A(0)) = \int_{-\infty}^0 d\tau ~K(t-\tau)~K(-\tau)~,
\label{mgndfga}
\ee
and
\be
{\rm E}[A(0)^2] =  \int_{-\infty}^0 d\tau ~[K(-\tau)]^2~.
\label{mgmsl}
\ee
For stationary processes such that $K(t)$ decays faster than $1/\sqrt{t}$ so 
as to make the integral in (\ref{mgmsl}) convergent, 
${\rm E}[A(0)^2]$ is a constant. We thus obtain the posterior ($t>0$) response 
(above the stationary average) to an 
endogenous shock occurring at time $t=0$ under the form of 
a conditional expectation of $A(t)$, conditioned by the existence of this
shock:
\be
{\rm E}_{\rm endo}[A(t)|A(0)=A_0] \propto A_0 \int_{0}^{+\infty} du ~K(t+u)~K(u)~,
\label{gjnwl}
\ee
for large $A_0$.
This relaxation of the activity after an endogeneous shock is in general 
significantly different from that given by (\ref{nmgjwslw}) following an exogeneous shock.

\subsubsection{Conditional noise trajectory}

What is the source of endogeneous shocks characterized by the response function
(\ref{gjnwl})?  To answer, let us consider the
process $W(t) \equiv \int_{-\infty}^t d\tau~{\hat \eta}(\tau)$, where ${\hat \eta}(t)
=\eta(t) - \langle \eta \rangle$ 
defines the centered innovations forcing the system (\ref{mbnall}).
Using the property (\ref{mgmms}), we find that for $t<0$
\be
\E_{\rm endo}[W(t)|A(0)=A_0] = \frac{\Cov[W(t),A(0)]}{\Var[A(0)]} \cdot
\left(A_0 - \E[A] \right) \propto \left(A_0 - \E[A] \right)
\int_{-\infty}^t d\tau ~K(-\tau)~,
\label{mgnvs}
\ee
where $\E_{\rm endo}[W(t)|A(0)=A_0] = 0$ for $t>0$ since the conditioning 
does not act after the shock.
Expression (\ref{mgnvs}) predicts that the expected path of the continuous
innovation flow prior to the endogeneous shock (i.e., for $t<0$)
grows like $\Delta W(t) = {\hat \eta}(t) \Delta t \sim K(-t) \Delta t$
upon the approach to the time $t=0$ of the large
endogeneous shock. In other words, conditioned on the
observation of a large endogeneous shock,
there is specific set of trajectories of the innovation flow $\eta(t)$ that led to
it. These conditional innovation flows have an expectation given by (\ref{mgnvs}).

Inserting the expression (\ref{mgnvs}) for the average conditional noise in
the definition (\ref{mbnall}) of the process, we obtain an expression proportional
to (\ref{gjnwl}). This shows
that the precursory activity preceding and announcing the endogeneous
shock follows the same time dependence as the relaxation (\ref{gjnwl})
following the shock, with the only modification that $t$ (for $t>0$ counting time after
the shock at $t=0$) is changed into $-t$ (for $t<0$ counting time before
the shock at $t=0$). We can also use (\ref{mgnvs}) into  (\ref{mbnall}) and
calculate the activity after the endogeneous shock to recover (\ref{gjnwl}).
These are two equivalent ways of arriving at the same result, the one using
(\ref{mgnvs}) illuminating the fundamental physical origin of the endogeneous
response.

These results allow us to understand
the distinctive features of an endogeneous shock compared to an
external shock. The later
is a single very strong instantaneous perturbation  that is sufficient in itself to
move the system
significantly according to (\ref{nbghlqa}). In contrast, an ``endogeneous'' shock
is the result of the cumulative effect of many small perturbations, each
one looking relatively
benign taken alone but, when taken all together collectively along the full
path of innovations, can add up coherently due to the long-range memory of the 
dynamical process. In summing, the term
``endogeneous'' is used here to refer to
the sum of the contribution of many ``small''
innovations adding up according to a specific most probable trajectory, as opposed
to the effect of a single massive external perturbation.

\subsection{Numerical simulation of an epidemic branching process with long-range 
memory}

To illustrate our predictions (\ref{nmgjwslw}) and (\ref{gjnwl}), we 
use a simple epidemic branching model defined as follows. The model 
describes the time evolution of the
rate of occurrence of events as a function of all past history. What is 
called ``event'' can be the creation of a new species or a new family of 
organisms as in \cite{CourtillotGaudemer}, the occurrence of an earthquake
as in \cite{SorSor,Helms}, the amplitude of the so-called financial volatility
as in \cite{Soretalfingerfin} or of aviation traffic, a change of weather regime, a climate
shift and so on. The rate $\lambda(t)$ of events at time $t$ is assumed
to be a function of all past events according to
\be
\lambda(t) =\sum_{i ~| ~t_i<t}~\phi(t-t_i)~,
\label{mgmzl}
\ee
where the sum is carried over all past events that occurred at times $t_i$
prior to the present $t$. The influence of such an event at a previous time $t_i$
is felt at time $t$ through the bare propagator $\phi(t-t_i)$. In our
present illustration, we consider a process equivalent to a fNm
with Hurst exponent $H=1/2+\theta$, which can
be shown to correspond to the choice $\phi(t) = \theta~c^{\theta}/(t+c)^{1+\theta}$,
where $c$ is an ultra-violet regularization time embodying 
a delay process at early times in the activity response after an event.
Indeed, the Master equation corresponding to the process (\ref{mgmzl})
can be shown \cite{SorSor,Helms} to be nothing but (\ref{mbnall}) with the renormalized
or dressed propagator $K(t) \propto 1/(t+c)^{1-\theta}$.

\begin{figure}
\begin{center}
\includegraphics[height=15cm]{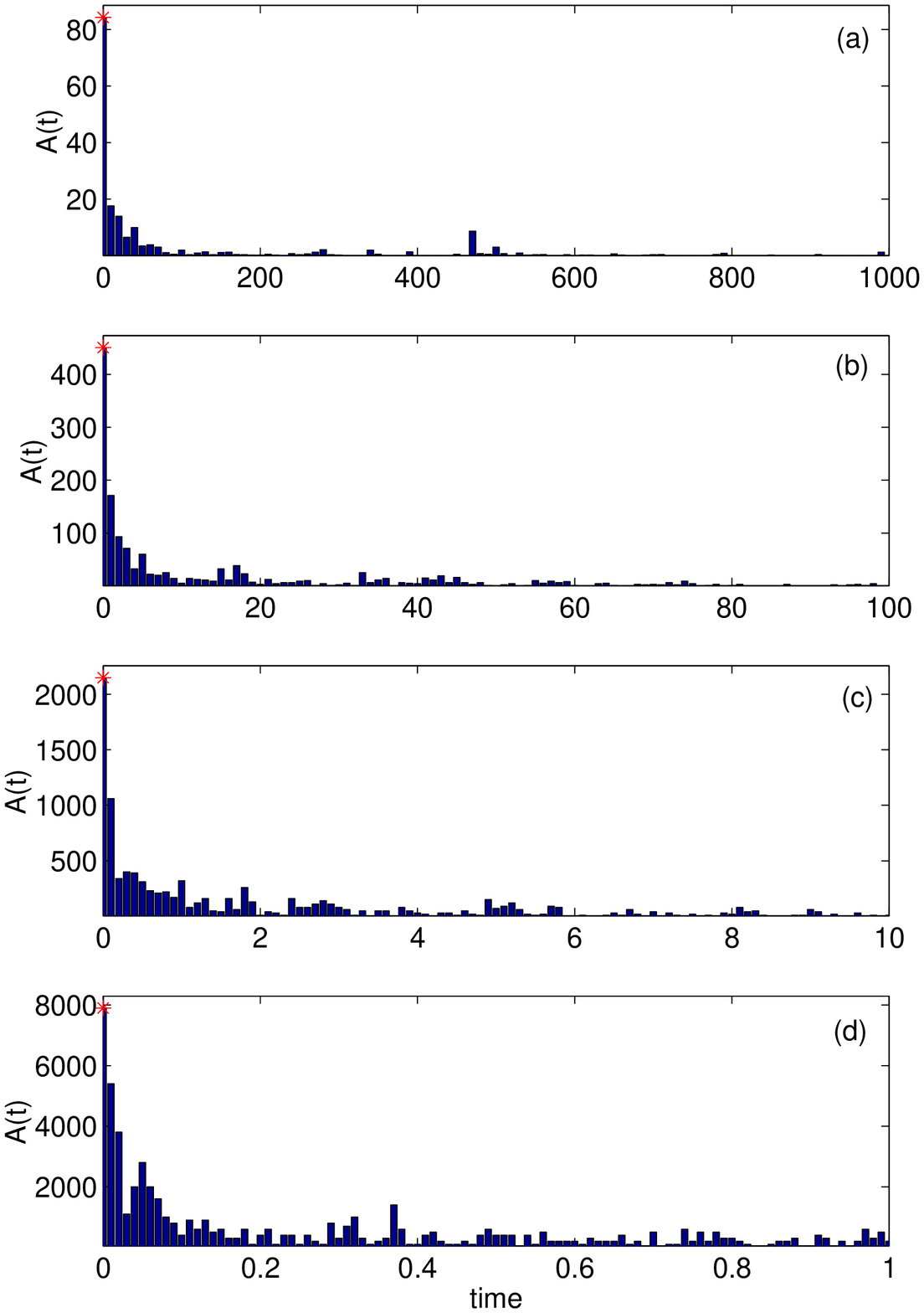}
\end{center}
\caption{Rate of activity following an exogeneous shock in a numerical simulation
of the epidemic branching model (\ref{mgmzl}) 
generated with a memory kernel decaying as a power law $\phi(t) \sim 1/(t+c)^{1+\theta}$
with parameters $\theta=0.1$ and $c=0.001$ without a
constant source term ($\mu=0$).
The rate of activity following an exogeneous shock that occurred at $t=0$  
is shown at increasing magnification from top to bottom. It
is evaluated using a bin size decreasing by factors of $10$ 
from  $\delta t=10$ (a) to  $\delta t=0.01$ (d). Averaging over many such
realizations would yield the average power law decay $K(t) \sim 1/(t+c)^{1-\theta}$
predicted by (\ref{nmgjwslw}).}
\label{Ntexolin} 
\end{figure}

\begin{figure}
\begin{center}
\includegraphics[height=15cm]{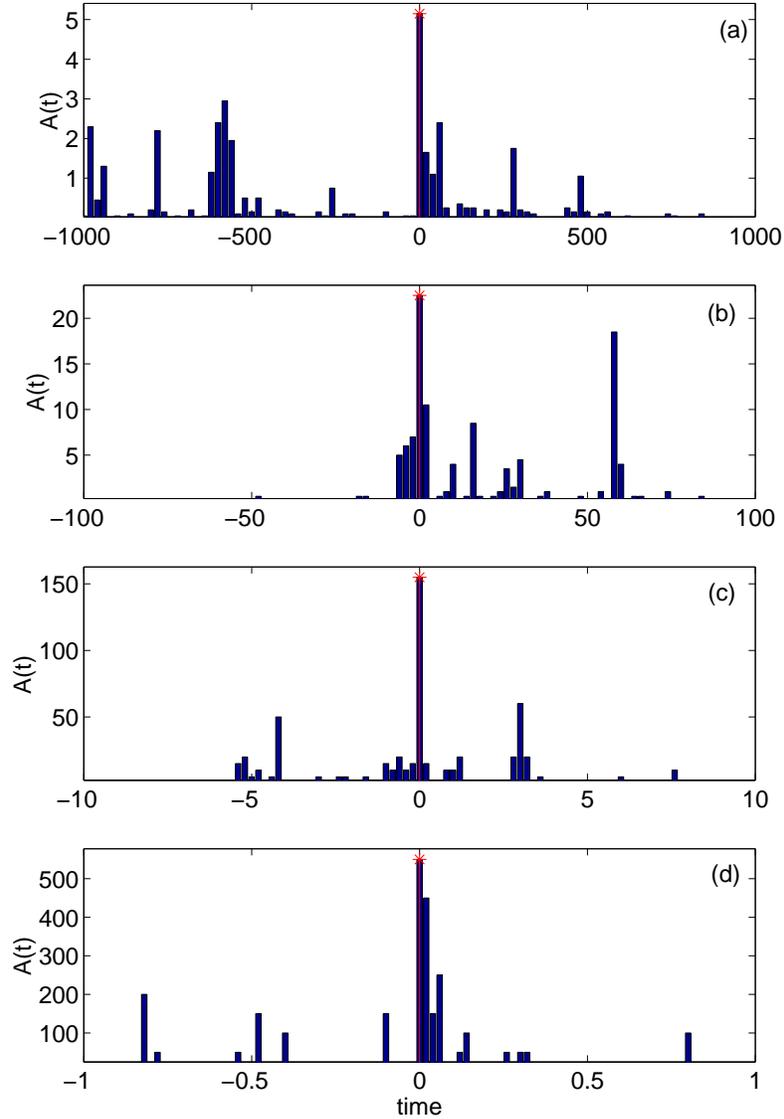}
\end{center}
\caption{Rate of activity prior to and after an endogeneously generated major burst of activity
generated by a numerical simulation of the epidemic branching model (\ref{mgmzl})
with a power law kernel with the same parameters 
$\theta=0.1$ and $c=0.001$ as in figure \ref{Ntexolin} with in 
addition a constant Poisson source term with rate $\mu=0.001$
corresponding, on average, to one event added from an external source
per $1000$ time units. Most of the observed activity is thus the result of
interactions between events.
The rate of activity close to the largest peak of activity is shown at 
increasing magnifications from top to bottom and is evaluated as in figure  \ref{Ntexolin}.
Both precursory and relaxational processes can be observed at many time scales.}
\label{Ntendolin}
\end{figure}

\begin{figure}
\begin{center}
\includegraphics[height=15cm]{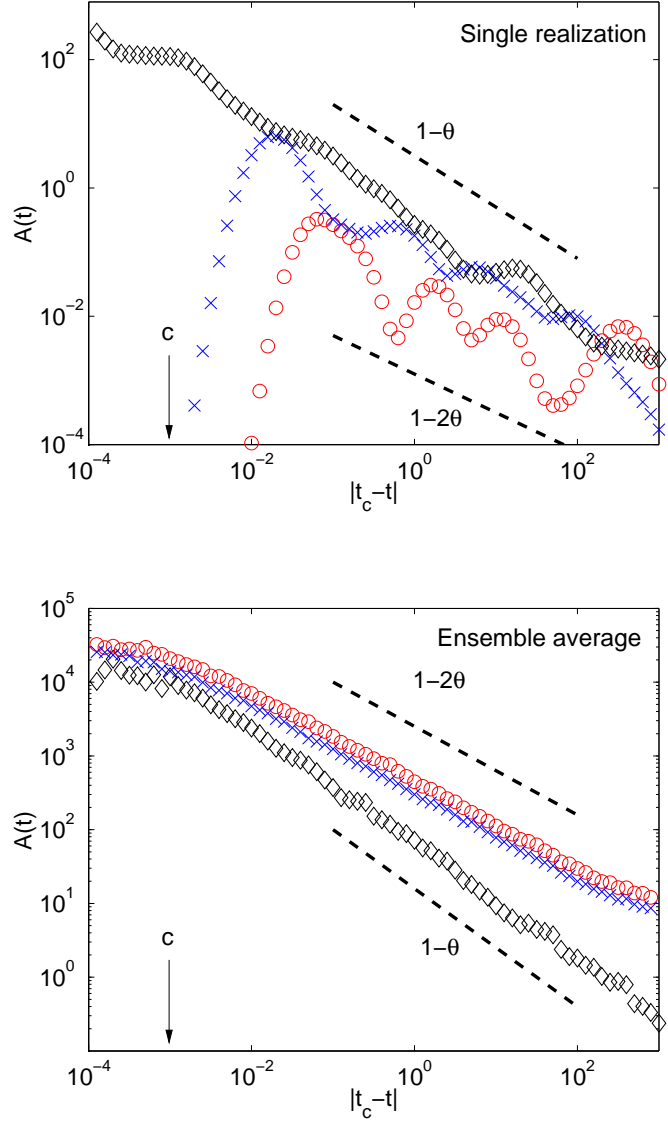}
\end{center}
\caption{Rate of activity for a single sequence (a) of the epidemic branching model
defined by (\ref{mgmzl}) generated with a memory kernel decaying as a power law
of time with the parameters $\theta=0.2$ and $c=10^{-3}$ and averaged over
many sequences (b).
The exogeneous relaxation is shown with diamonds, the endogeneous relaxation
is shown as crosses and circles are for the precursory activity in the endogeneous case.
Large fluctuations are observed in the precursory activity and in the endogeneous relaxation
when looking at a single sequence, due to the small number $\approx
100$ of observed events. Averaging over 50 realizations,
we see clearly the faster decay rate $\sim 1/t^{1-\theta}$ for the exogeneous
relaxation predicted by (\ref{nmgjwslw})
compared with the endogenous one predicted by (\ref{gjnwl}). The same decay rate $\sim
1/t^{1-2\theta}$ predicted by (\ref{gjnwl}) is observed for both the endogeneous 
precursory and post-event relaxation.}
\label{endoexo}
\end{figure}

Numerical simulations of the epidemic branching process are performed by drawing
events in succession according to a non-stationary Poisson process
with instantaneous rate (\ref{mgmzl}).
Figures \ref{Ntexolin} and \ref{Ntendolin} 
show successive magnifications of 
time series of the activity rate after an exogeneous shock and around
an endogeneous shock, respectively, in order to visualize
the precursory and relaxation activities. In figure \ref{Ntendolin},
an external source of activity necessary for seeding
has been added as a Poisson process of rate
$\mu = 10^{-3}$ corresponding on average to one external 
event over a time interval of $1000$.
The most striking visual difference is
the existence of the precursory signal occurring at many time scales for the
endogeneous shock.
Figure \ref{endoexo} quantifies the precursory and relaxation rates
associated with activity shocks.
The top panel shows the relaxation of the activity (rate of events) 
following an external shock
compared to that after an endogeneous shock, for a single realization.
$t_c$ is the time of the shock. The horizontal axis is $t-t_c$ for
the relaxation of the activity after the shock. The precursory activity prior 
to the shock is also shown for the endogeneous shock as a function of $t_c-t$.
The bottom panel shows the same three activity functions after averaging over
many realizations, translating time in the averaging so that all shocks occur at the same
time denoted $t_c$. The prediction (\ref{nmgjwslw}) states that the relaxation of the
activity after an exogeneous shock should decay as $K(t) \propto 1/(t-t_c+c)^{1-\theta}$
while the decay after an endogeneous shock should be given by (\ref{gjnwl})
which predicts the law $\propto 1/(t-t_c+c)^{1-2\theta}$, that is, a significantly smaller
exponent for $\theta > 0$. Similarly, we predict that the precursory activity prior
to an endogeneous shock should increase as $\propto 1/(t_c-t+c)^{1-2\theta}$. These
predictions are verified with very good accuracy, as seen in figure  \ref{endoexo}.

These simulations confirm that there is a distinctive difference in the relaxation
after an endogeneous shock compared to an exogeneous shock, if the memory kernel
is sufficiently long-ranged. For
a single realization, there are unavoidable fluctuations that may blur out this
difference. However, we see a quite visible precursory signal (foreshock activity)
that is symmetric to that relaxation process in the case of an endogeneous shock.
This follows from the model used here which obeys the time-reversal symmetry.
This may be used as a distinguishing signature of an endogeneous shock.

\section{Classification of the distinctive responses for
different classes of memory kernels}

The family of power law kernels used in the simulations presented in
figures \ref{Ntexolin}, \ref{Ntendolin} and \ref{endoexo} are only one possibility
among many. Our formalism allows us to classify the distinctive properties
of the relaxation and precursory behaviors that can be expected for
an arbitrary memory kernel. We now provide this classification by 
studying (\ref{nmgjwslw}) and (\ref{gjnwl}).

\subsection{Short-time response}

We compare the initial slopes of the relaxations 
after the occurrence of the shock at $t=0$. Thus, by
short-time, we mean the asymptotic decay law just after the shock.
For this, we expand (\ref{nmgjwslw}) to get
\be
{\rm E}_{\rm exo}[A(t)] =  A_0 ~K(0) \left[1 + {K'(0) \over K(0)} ~~t + {\cal O}(t^2)\right]
=  A_0 ~K(0) \left[1 + {d \ln K \over dt}|_{t=0} ~~t + {\cal O}(t^2)\right]~,
\label{nmgjwsaalw}
\ee
where $K'(t)$ denotes the derivative of $K(t)$ with respect to time.

Similarly, expanding the integral in (\ref{gjnwl}) for short times, we obtain
\be
{\rm E}_{\rm endo}[A(t)|A(0)=A_0] \propto A_0 F(0) 
\left[ 1 + {1 \over 2}{d \ln F \over dt}|_{t=0} ~~t + {\cal O}(t^2)\right]~,
\label{gjnwssl}
\ee
where
\be
F(t) \equiv \int_t^{+\infty} du [K(u)]^2~
\ee
is a monotonically decreasing function of time.

It is convenient to use the parameterization
\be
F(t) = e^{-g(t)}~,
\label{gnbbgnlw}
\ee
where $g(t)$ is an monotonously increasing function of time. Inserting
(\ref{gnbbgnlw}) in (\ref{nmgjwsaalw}) and (\ref{gjnwssl}) leads to
\be
{\rm E}_{\rm exo}[A(t)] = A_0 ~K(0) \left[ 1 - \left({1 \over 2} g'(0) -{1\over 2}
{g''(0) \over g'(0)}\right) ~t + {\cal O}(t^2) \right]~,
\ee
and
\be
{\rm E}_{\rm endo}[A(t)|A(0)=A_0] \propto A_0 F(0) 
\left[ 1 - {1 \over 2} g'(0) ~t + {\cal O}(t^2) \right]~.
\ee

\begin{enumerate}
\item For $g''(0)=0$, that is, $g(t) = 2\alpha t$ corresponding 
to an pure exponential relaxation $K(t) \propto \exp [-\alpha t]$, 
the velocities of the responses to an exogeneous and to endogeneous shock
are identical;

\item for $g''(0) >0$ corresponding to a super-exponential relaxation
$K(t) \propto \exp [-\alpha t^c]$ with $c>1$, the exogeneous relaxation
is slower than the endogeneous one;

\item for $g''(0) <0$ corresponding to a sub-exponential relaxation
such as a stretched exponential
$K(t) \propto \exp [-\alpha t^c]$ with $c<1$ or to 
the family of regularly varying functions such as power laws, the exogeneous relaxation
is faster than the endogeneous one.
\end{enumerate}
The exponential relaxation thus marks the boundary between two opposite regimes.
As is intuitive, a sub-exponential relaxation betraying a long memory process
leads to a slower short-time recovery after an endogeneous shock, because
it results from a long preparation process (\ref{mgnvs}).

\subsection{Asymptotic long-time response}

Since $K(t)$ is a monotonously decaying function, $K(t+u) \leq K(t)$ for any $u \geq 0$.
This leads to the following inequality
\be
{\rm E}_{\rm endo}[A(t)|A(0)=A_0] \leq A_0 K(t) \int_{0}^{+\infty} du~ K(u)~,
\label{gjnwaal}
\ee
which is valid if the integral $\int_{0}^{+\infty} du ~K(u)$ exists, that is, if
$K(t)$ decays faster than $1/t$ at large times. This shows that,
as soon as $K(t) \ll C/t$ for any positive constant $C$, 
${\rm E}_{\rm endo}[A(t)|A(0)=A_0] < {\rm E}_{\rm exo}[A(t)]$. But the 
difference may be small and inobservable. 
For instance, for $K(t) \propto 1/t^{1+\theta}$ with $\theta > 0$,
a careful examination of the integral in (\ref{gjnwl}) shows that, due to 
the contribution of the conditional noise close to the shock, we have
\be
{\rm E}_{\rm endo}[A(t)|A(0)=A_0] \propto {A_0 \over t^{1+ \theta}}
\sim  {\rm E}_{\rm exo}[A(t)] ~.
\label{gwgjw}
\ee
Thus, there is no qualitative difference 
in the relaxation rates of an endogeneous shock and exogeneous shock in this case:
the contributions of all the conditional activity prior to the
endogeneous shock is equivalent to that of the shock itself. A more elaborate and
analysis specific to the problem at hand 
must be performed to predict the prefactors that will be 
different in the endogeneous and exogeneous cases.

In constrast, for memory kernels $K(t) \propto 1/t^{1-\theta}$ with $\theta > 0$
decaying slower than $1/t$, 
as for a stationary fNm of the form (\ref{ngjlk}), we obtain
\be
{\rm E}_{\rm endo}[A(t)|A(0)=A_0] \propto {A_0 \over t^{1-2 \theta}}
\gg  {\rm E}_{\rm exo}[A(t)] \propto  {A_0 \over t^{1-\theta}}~.
\label{gwwwggw}
\ee
In this case, the relaxation following an endogeneous shock decays significantly more slowly
than for an exogeneous shock. This case is examplified in figure \ref{endoexo}. 
In the long time limit, the decay law $1/t$ thus marks the boundary between
two opposite regimes.

\subsection{Illustration}

An illustration of this critical behavior is provided by the response of the 
price volatility $\sigma_{\Dt}$ at scale $\Dt$
defined as the amplitude (absolute value) of the return
 $r_{Dt}(t) \equiv \ln [p(t)/p(t-\Dt)] = \epsilon(t) \cdot 
\sigma_{\Dt}(t) = \epsilon(t) \cdot e^{\w_{\Dt}(t)}$. 
of a financial asset. $\epsilon(t)$ is a random sign.
Indeed, financial price time series have been shown to exhibit a long-range
correlation of their log-volatility $\w_{\Dt}$, described by 
a model \cite{M_etal,Soretalfingerfin} in 
which $\w_{\Dt}(t)$ follows the process (\ref{mbnall})
with 
\be
K_\Dt(t) \sim K_0 \sqrt{\frac{\lambda^2 T}{t}}~~~~ {\rm for}~~\Dt \ll t \ll T~,
\label{ngbjrk}
\ee
where $T \approx 1$ year is a so-called integral time scale. This form (\ref{ngbjrk})
corresponds to the parameterization (\ref{ngjlk}) with $\theta =1/2$.
Sornette et al. \cite{Soretalfingerfin} have shown that 
there is a clear distinction between the relaxation of stock market
volatility after an exogeneous event such at the September 11, 2001 attack or the
Aug., 19, 1991 coup against Gorbachev and that after an endogeneous event such 
as the October 19, 1987 crash. In this model, the long-range memory acting
on the logarithm of the volatility induces an additional effect, namely the
exponent of the power law relaxation after an endogeneous shock is a linear
function of the amplitude of the shock.

\subsection{Synthesis of the asymptotic short- and long-time regimes}

We have found two special functional forms for the response kernel $K(t) 
\propto 1/t$ and $K(t) \propto \exp [-\alpha t]$, which are ``invariant''
or indifferent
with respect to the endogeneous versus exogeneous origin of a shock. 
Thus, for normal exponential relaxation processes as well as 
for power relaxation $\propto 1/t$, the functional form
of the recovery does not allow one to distinguish between an endogeneous
and an exogeneous shock.

These two invariants $K(t) \propto \exp [-\alpha t]$ and $K(t) 
\propto 1/t$ delineate two opposite regimes, the first one for short-time scales
and the second one for long-time scales:
\begin{enumerate}
\item for $K(t) \propto 1/t^{1-\theta}$ with $\theta > 0$, the endogeneous 
response decays more slowly than the exogeneous response, at all time scales;

\item for $\exp [-\alpha t] \ll K(t) \ll 1/t$ for any positive $\alpha$, the endogeneous 
response decays more slowly than the exogeneous response at short time scales
and has the same dependence as the exogeneous response at long time scales; this regime
describes for instance the stretched exponential relaxation of complex fluids 
alluded to above;

\item for $K(t) \ll \exp [-\alpha t] $ for any positive $\alpha$, the endogeneous 
response decay faster than the exogeneous response at all time scales.
\end{enumerate}

More complicated behaviors can occur when the memory kernel 
$K(t)$ exhibits a change of regime, crossing the exponential and/or $1/t$
boundaries at certain time scales. Each situation requires a specific analysis
which yields sometimes surprising non-intuitive results \cite{HelmsinvOmori}.

\section{Conclusion}

We think that the conceptual framework presented here may be applied
to a large variety of situations, beyond those alluded to in the introduction.
For instance, the result (\ref{mgnvs}) has been shown 
to explain the so-called inverse Omori's law for
earthquake foreshock activity before a mainshock, in a simple model of
earthquake triggering \cite{HelmsinvOmori}. The same mechanism may explain
the premonitory seismicity pattern known as ``burst of aftershocks'' 
\cite{kb78}: a mainshock with an abnormally large number of aftershocks
has been found to be a statistically significant precursor to strong earthquakes
\cite{Molchan}.

Many dynamical systems in Nature, such as geophysical
and biological systems (immune network, memory processes
in the brain, etc.), or created by man such as social structures and networks
(Internet), States and so on, exhibit long-memory effects due to a 
wealth of possible mechanisms. For instance, Krishan Khurana
at UCLA has suggested to us that 
the concept proposed here could explain
that endogeneous civil wars have long-lasting effects with slow reconstruction
compared with the fast recovery after exogeneous wars (that is, imposed or coming
from the outside).
The increasing emphasis on the concepts of emergence and complexity has
emphasized an endogeneous origin of the complicated dynamical behavior of
complex systems. In reality, most (so-called) complex systems are the result of their internal
dynamics/adaptation in response to a flow of external perturbations, but some
of these external perturbations are rare extreme shocks. What is the role of these
exogeneous shocks in the self-organization of a complex system? Can one 
distinguish the impact of extreme exogeneous shocks from an endogeneous organization
at different time scales? Our present analysis has just scratched the surface of these
important and deep questions by suggesting an angle of attack based
on the conditional historical process at the basis of strong endogeneous 
fluctuations. Extensions of the present simplified framework involve the
generalization to multidimensional coupled processes such as in 
\cite{Johnson} and to nonlinear spatio-temporal processes.

We are grateful to A.B. Davis. V. Keilis-Borok and V.F. Pisarenko for useful exchanges.
This work was partially supported by
the James S. Mc Donnell Foundation 21st century scientist award/studying
complex system.

\pagebreak

\newpage

\pagebreak

\clearpage
\pagebreak

\clearpage
\pagebreak

\end{document}